\documentclass[pdflatex,sn-mathphys-num]{sn-jnl}

\usepackage{amsmath,amssymb,amsfonts}%

\usepackage{amsthm}%
\usepackage[title]{appendix}%
\usepackage{booktabs}%
\usepackage{braket}%
\usepackage{chemformula}%
\usepackage{fancyhdr}%
\usepackage[symbol]{footmisc}
\usepackage{graphicx}%
\usepackage{hyperref}%
\usepackage{mathrsfs}%
\usepackage{multirow}%
\usepackage{placeins}%
\usepackage{siunitx}%
\usepackage{svg}%
\usepackage{textcomp}%
\usepackage{xcolor}%

\usepackage[capitalise,noabbrev]{cleveref}%

\makeatletter
\newcommand{\@aionnum}{xxxxxx}
\newcommand{\aionnum}[1]{%
    \def\@aionnum{#1}%
}
\makeatother

\usepackage{silence}
\makeatletter
\def\@xfootnote[#1]{%
  \protected@xdef\@thefnmark{#1}%
  \@footnotemark\@footnotetext}
\makeatother

\usepackage{globalvals}
\usepackage{siunitx}

\defVal{median-atoms-top}{\num{3100(210)}}
\defVal{median-atoms-bottom}{\num{2040(160)}}
\defVal{peak-dev-percent}{15\%}

\defVal{drift-top}{-42}
\defVal{drift-bottom}{8}

\defVal{atom-number-sigma-rel}{8}  

\defVal{qpn-contrast-top}{0.81}
\defVal{qpn-contrast-bottom}{0.84}

\defVal{num-phase-steps}{100}
\defVal{stddev-random-phase}{$4\pi\,\si{rad}$}
\defVal{stark-pulse-detuning}{\qty{-80}{MHz}}
\newcommand\piPulseDuration{\qty{44}{\micro\second}}

\newcommand\lowLaserNoiseSigmaOneBlock{\qty{3.69(19)}{\milli\radian}}
\newcommand\highLaserNoiseSigmaOneBlock{\qty{3.89(20)}{\milli\radian}}
\newcommand\lowLaserNoiseSigmaWholeDataset{\qty{260(13)}{\micro\radian}}
\newcommand\highLaserNoiseSigmaWholeDataset{\qty{275(14)}{\micro\radian}}
\newcommand\highLaserNoiseSigmaMinusSQL{\qty{16(17)}{\micro\radian}}
\newcommand\lowLaserNoiseSigmaMinusSQL{\qty{2(16)}{\micro\radian}}
\newcommand\differenceSigmaWholeDataset{\qty{14(19)}{\micro\radian}}
\newcommand\dataNumShots{\num{56623}}
\newcommand\dataNumShotsSingle{\num{28312}}
\newcommand\dataDuration{\qty{61.9}{h}}
\newcommand\numMCSims{\num{5100}}
\newcommand\shotsPerBin{\num{141}}

\def\excitationFractionTop{\qty{20}{\percent}}
\def\excitationFractionBottom{\qty{62}{\percent}}

\def\sequenceTime{\qty{200}{\micro\second}}

\newcommand\theoreticalSQLSingleShot{\qty{43.5(16)}{\milli\radian}}

\newcommand\theoreticalSQLWholeDataset{\qty{258(10)}{\micro\radian}}

\def\dStark{\ensuremath{\phi_\mathrm{Stark}}}

\def\dPhi{\ensuremath{\delta\phi}}

\newcommand\injectedSignalAmplitudeForFreqs{\qty{17.3}{\milli\radian}}

\DeclareMathOperator{\atantwo}{atan2}

\begin{document}

\aionnum{2026-03}
\title{\LARGE
  \textbf{A Prototype Differential Atom Interferometer for Fundamental Physics}

  \vspace{0.5em}
}

\author[1]
{C.~F.~A.~Baynham\footnote{\label{equal}Lead authors who contributed equally to this work}\ , R.~Hobson\footnotemark[\value{footnote}]\ , O.~Buchmueller (O.~Buchm{\"u}ller), D.~Evans, L.~Hawkins, L.~Iannizzotto~Venezze, A.~Josset, D.~Lee, E.~Pasatembou,
B.~E.~Sauer, M.~R.~Tarbutt, T.~Walker;}

\author[2]{\vspace{2mm}\\O.~Ennis, U.~Chauhan, A.~Brzakalik, S.~Dey, S.~Hedges\footnote{Present address: Nomad Atomics, 33 Elizabeth Street, Richmond, Victoria 3121, Australia}\ , B.~Stray\footnote{Present address: Jet Propulsion Laboratory, California Institute of Technology, Pasadena, California 91109, USA}\ , M.~Langlois\footnote{Present address: Jet Propulsion Laboratory, California Institute of Technology, Pasadena, California 91109, USA}\ , K.~Bongs\footnote{Present address: Institute of Quantum Technologies, German Aerospace Center (DLR), Wilhelm-Runge-Stra{\ss}e 10, 89081 Ulm, Germany}\ , T.~Hird, S.~Lellouch, M.~Holynski;}

\author[3]{\vspace{2mm}\\B.~Bostwick\footnote{Present address: Kirchhoff-Institut f{\" u}r Physik, Universität Heidelberg, Im Neuenheimer Feld 227, 69120 Heidelberg, Germany}\ , J.~Chen, Z.~Eyler, V.~Gibson, T.~L.~Harte, C.~C.~Hsu, M.~Karzazi, C.~Lu, B.~Millward, J.~Mitchell, N.~Mouelle, B.~Panchumarthi~\footnote{Present address: Department of Physics \& Astronomy, Northwestern University, 2145 Sheridan Road, Evanston, Illinois 60208-3112, United States}\ , J.~Scheper, U.~Schneider, X.~Su\footnote{Present address: Physikalisches Institut, Universit{\" a}t Tübingen, Auf der Morgenstelle 14, 72076 T{\" u}bingen, Germany}\ , Y.~Tang, K.~Tkalcec (K.~Tkal{\v c}ec), M.~Zeuner\footnote{Present address: Ludwig-Maximilians-Universit{\" a}t München, Geschwister-Scholl-Platz 1, 80539 München, Germany}\ , S.~Zhang, Y.~Zhi\footnote{Present address: Department of Physics, University of Virginia, 382 McCormick Road, Charlottesville, VA 22904, USA}\ ;}

\author[4]{\vspace{2mm}\\L.~Badurina\footnote{Present address: Walter Burke Institute for Theoretical Physics, California Institute of Technology, Pasadena, CA 91125,
USA}\ , A.~Beniwal\footnote{Present address: Intersect Australia, Sydney, Australia}\ , D.~Blas\footnote{Present address: Institut de F\'{i}sica d’Altes Energies (IFAE), The Barcelona Institute of Science and Technology,
Campus UAB, 08193 Bellaterra (Barcelona), Spain  and Instituci\'{o} Catalana de Recerca i Estudis Avan\c{c}ats (ICREA), Passeig Llu\'{i}s Companys 23, 08010 Barcelona, Spain}\ , J.~Carlton, J.~Ellis, C.~McCabe, G.~Parish, D.~Pathak~Govardhan, V.~Vaskonen\footnote{Present address: Keemilise ja Bioloogilise F\"{u}\"{u}sika Instituut, R\"{a}vala pst. 10, 10143 Tallinn, Estonia}\ \ ;}

\author[5]{\vspace{2mm}\\T.~Bowcock, K.~Bridges, A.~Carroll, J.~Coleman, G.~Elertas, S.~Hindley, C.~Metelko, H.~Throssell, J.~N.~Tinsley;}

\author[6]{\vspace{2mm}\\E.~Bentine, M.~Booth, D.~Bortoletto, N.~Callaghan, C.~Foot, C.~Gomez-Monedero (C.~G{\'o}mez-Monedero), K.~Hughes, A.~James, T.~Leese, A.~Lowe, J.~March-Russell, J.~Sander, J.~Schelfhout, I.~Shipsey\footnote[$\dagger$]{Deceased}\ , D.~Weatherill, D.~Wood;}

\author[7]{\vspace{2mm}\\S.N.~Balashov, M.G.~Bason,}\author[5,7]{K.~Hussain,}\author[7]{H.~Labiad, P.~Majewski, A.L.~Marchant, D.~Newbold, Z.~Pan, Z.~Tam, T.C.~Thornton, T.~Valenzuela, M.G.D.~van der Grinten, I.~Wilmut;}

\author[8]{\vspace{2mm}\\K.~Clarke, A.~Vick}

\affil[1]{Department of Physics, Blackett Laboratory, Imperial College London, Prince Consort Road, London, SW7 2AZ, UK}

\affil[2]{Physics and Astronomy, University of Birmingham, Edgbaston, Birmingham, B15 2TT, UK}

\affil[3]{Cavendish Laboratory, J J Thomson Avenue, University of Cambridge, Cambridge, CB3 0HE, UK}

\affil[4]{Physics Department, King's College London, Strand, London, WC2R~2LS, UK}

\affil[5]{Department of Physics, University of Liverpool, Merseyside, L69 7ZE, UK}

\affil[6]{Department of Physics, University of Oxford, Parks Road, Oxford, OX1~3PU, UK}

\affil[7]{Rutherford Appleton Laboratory, UKRI-STFC, Harwell Campus, Didcot, OX11 OQX, UK}

\affil[8]{STFC Daresbury Laboratory, Warrington, WA4 4AD, UK}

\date{\today}


\abstract{%
Gravitational waves and ultralight dark matter are among the most compelling frontiers in fundamental physics, motivating proposals for Very Long-Baseline Atom Interferometers (VLBAIs) such as AION~\cite{Badurina2020}, MAGIS~\cite{Abe2021}, AICE~\cite{baynham_letter_2025} and AEDGE~\cite{AEDGE:2019nxb} that aim to detect frequencies at which ground-based~\cite{LIGOScientific:2014pky} and space-borne~\cite{LISA:2017pwj} laser interferometers lose sensitivity. VLBAIs look for signals by comparing the quantum phase evolution of widely separated atomic ensembles interrogated by a common laser. However, their performance depends critically on suppressing noise sources, particularly laser phase noise. Experimental validation of such noise rejection remains an important challenge.

Here we demonstrate a prototype differential atom interferometer based on the single-photon clock transition of fermionic \ch{^{87}Sr}, realising for the first time a gradiometer configuration with a species intrinsically suited to kilometre-scale and space-baseline operation. The instrument operates at the Standard Quantum Limit~\cite{pezze_quantum_2018} with no excess noise beyond atom shot noise, and the differential configuration maintains quantum-limited sensitivity in the presence of several radians of artificially injected laser phase noise per shot, emulating the conditions expected in a VLBAI. We further demonstrate recovery of coherent oscillatory signals across a broad frequency range under fully phase-randomised conditions, a capability that is inaccessible to a single interferometer operating in the same regime.

These results provide an experimental validation of the noise-immune measurement principle underlying VLBAIs and mark an important step towards next-generation quantum sensors for gravitational-wave detection and searches for ultralight dark matter~\cite{Arvanitaki:2016fyj,Badurina:2021rgt}.

\vspace{5mm}\textit{Version published in Nature volume 654, pages 622-628 (2026)}
}

\pagestyle{fancy}
\maketitle
\pagestyle{fancy}

The discovery of gravitational waves (GWs) by the LIGO and Virgo laser interferometer experiments~\cite{LIGOScientific:2016aoc} has opened a new window on the Universe, with prospects for breakthroughs in fundamental physics, astrophysics and cosmology. Just as observations of electromagnetic waves over a wide range of frequencies have provided insights into physical processes within and beyond our galaxy, as well as in the primordial Universe, it is expected that GW observations over a wide range of frequencies will offer complementary insights into an equally rich spectrum of phenomena. The operating terrestrial laser interferometer detectors, LIGO, Virgo and KAGRA, are sensitive to GWs at frequencies around \qtyrange{e1}{e3}{\hertz}~\cite{LIGOScientific:2014pky,VIRGO:2014yos,Aso:2013eba}, and the Laser Interferometer Space Antenna (LISA) experiment, now under construction, will be most sensitive to GWs with frequencies around \qtyrange{e-4}{e-1}{\hertz}~\cite{LISA:2017pwj}, leaving unexplored an intermediate range of frequencies around \qtyrange{e-1}{e1}{\hertz}.

Important sources of GWs in this frequency range are mergers of intermediate-mass black holes, heavier than those detected by ground-based laser interferometers, and lighter than those targeted by LISA.\@ Such intermediate-mass black holes are thought to provide building blocks for the supermassive black holes~\cite{Ellis:2024nzv} at the hearts of most galaxies, so measurements of their mergers using long-baseline atom interferometers~\cite{Proceedings:2023mkp,Proceedings:2024foy} could reveal how supermassive black holes are formed~\cite{Ellis:2023iyb}. Further, observations of the slowly evolving inspiral stages of solar-mass mergers would be possible for days or weeks instead of seconds,
enabling multi-messenger astronomy by pinpointing the locations of GW sources in the sky~\cite{baum_gravitational_2024}.

\begin{figure}[tb]
	\centering
    \includegraphics[width=0.7\textwidth]{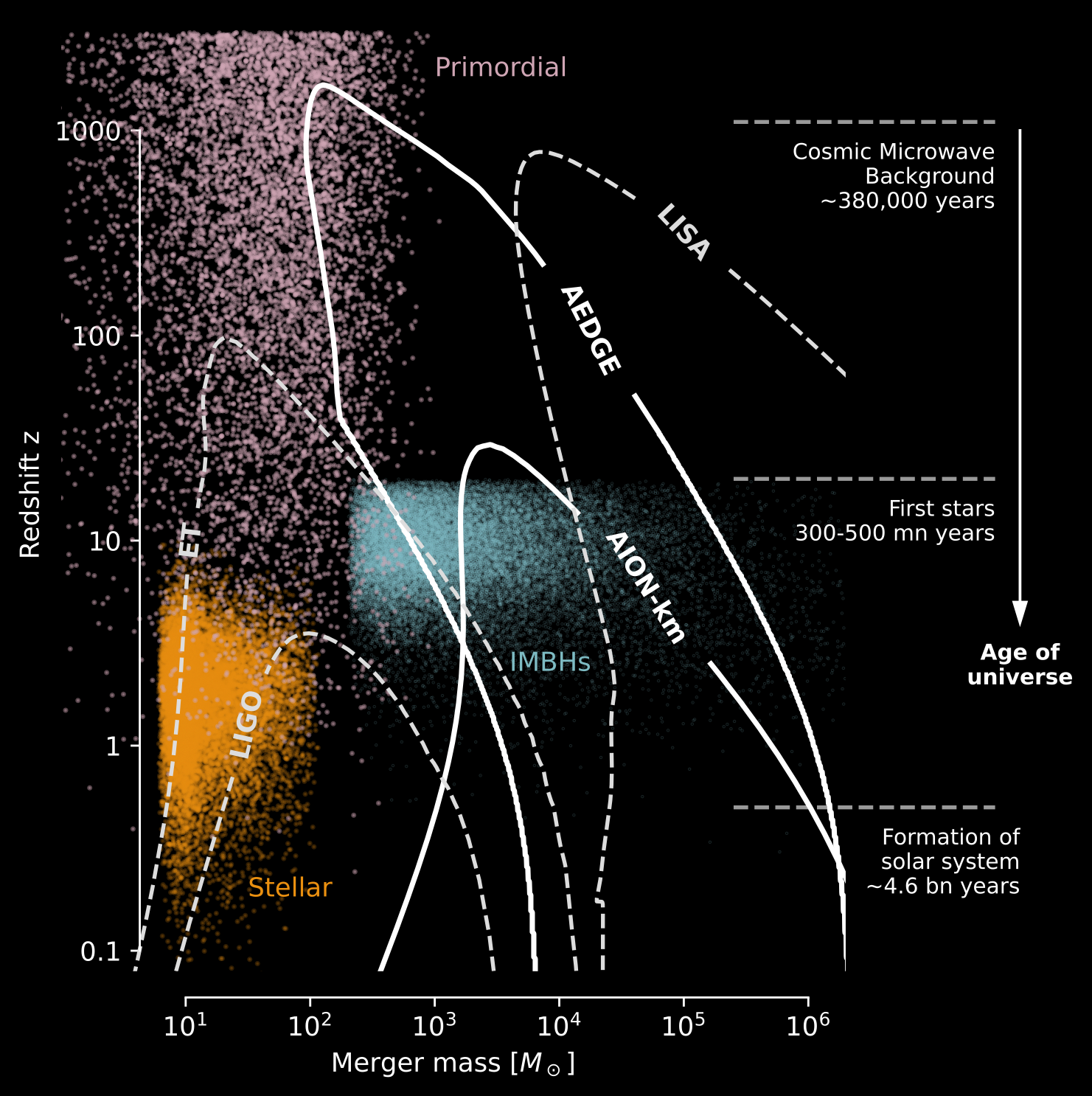}
\caption{%
\textbf{The parameter space of black hole mergers probed by various GW detectors, both operational and planned.} The horizontal axis gives the mass $M$ of the black hole merger causing the GW, in units of the solar mass. The vertical axis is the distance to the GW source, expressed as the redshift $z$.
The cyan dots are GW signals from a simulation of a one-year data sample of black hole mergers generated using a hierarchical model of the formation of supermassive black holes~\cite{Ellis:2024nzv}, the orange dots are GW signals from a simulated sample of stellar mass black hole mergers, and the violet dots are GW signals from a hypothetical population of primordial black holes (see Methods for details). Also shown are the prospective sensitivities of different detectors, including laser-interferometer detectors~\cite{LIGOScientific:2014pky,LISA:2017pwj,evans_cosmic_2023} as well as AION-km~\cite{Badurina:2021rgt} and AEDGE~\cite{AEDGE:2019nxb} atom-interferometer detectors with baselines of \qty{1}{\kilo\meter} and \qty{40000}{\kilo\meter} respectively. This figure was inspired by the Cosmic Explorer proposal~\cite{evans_cosmic_2023}.
}%
\label{fig:gravitational-waves}
\end{figure}

\begin{figure}[tb]
	\centering
    \includegraphics[width=\textwidth]{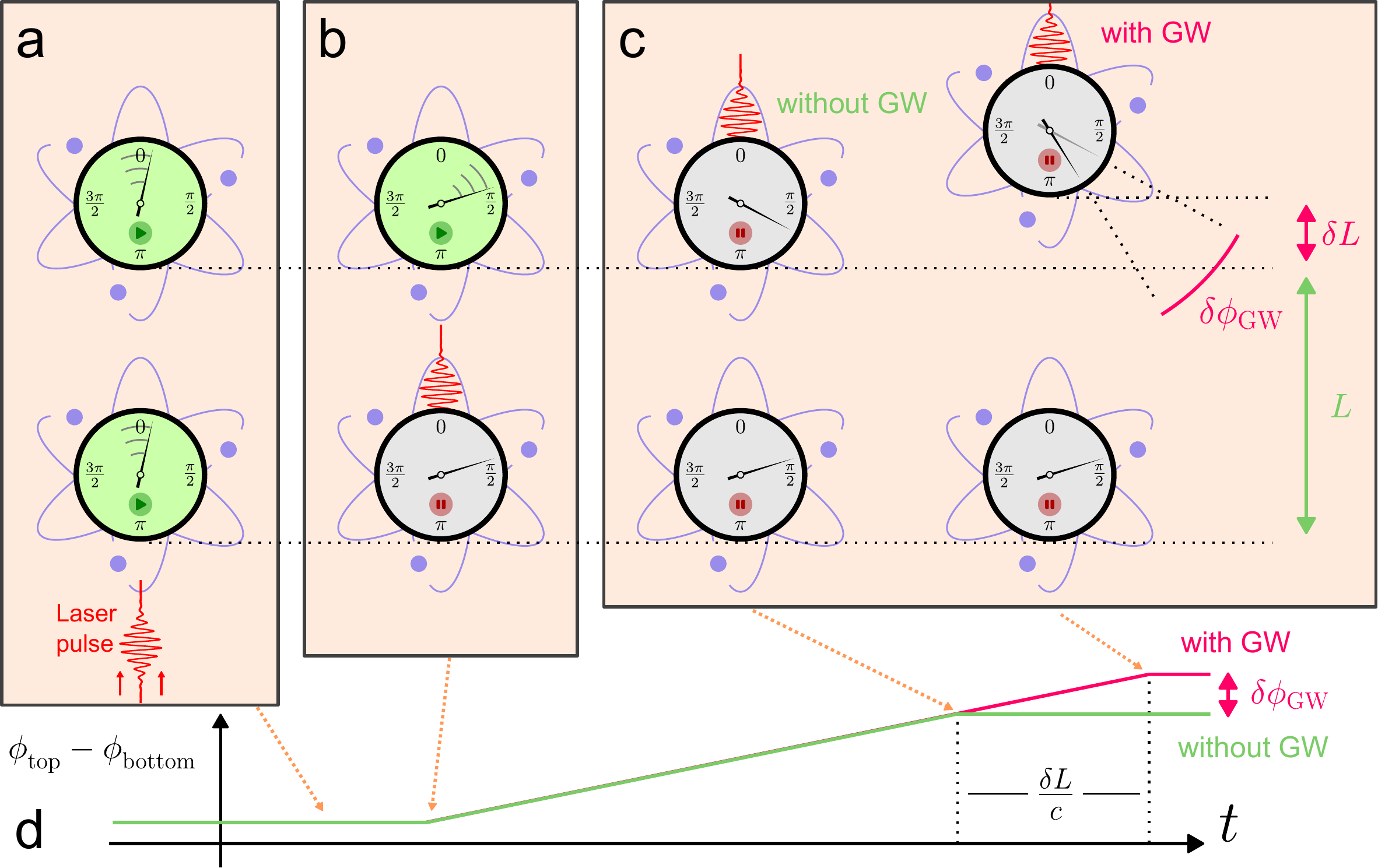}
\caption{%
    \textbf{An illustration of the detector's sensitivity to gravitational waves.} %
    In the moments before the final $\pi/2$ beam splitter pulse (see \cref{fig:experimental-overview}), the two atom interferometers can be treated as freely-falling atomic clocks \textit{(a)} accruing phase at a rate %
    $\omega_0$. The pulse halts this accrual of phase for the lower cloud, resulting in an accrual of differential phase \textit{(b, d)} that continues until the pulse reaches the second cloud \textit{(c)}. %
    In the proper frame of the bottom cloud (as pictured), the atoms are displaced by a transient gravitational wave. This has the effect of delaying (or hastening) this second interaction, imparting (at leading order) a detectable differential phase of %
    $\dPhi_\text{GW} = \pm \frac{\delta L}{c} \omega_0$~\cite{badurina2024signatureslinearizedgravityatom}. %
    Crucially, any phase noise due to the laser pulse itself is strongly suppressed in the differential measurement since it impacts both interferometers equally. The mechanism for sensitivity to dark matter (not pictured) is similar, but results from modulation of $\omega_0$ instead of $L$---see Refs.~\cite{Graham:2015ifn,
    Arvanitaki:2016fyj,Badurina:2021lwr,Badurina:2023wpk,Blas:2024jyh} for descriptions. %
    This simplified picture neglects complications arising from other interferometer phases $\phi_\text{other}$~\cite{Abe2021}, the other pulses in the sequence~\cite{graham_new_2013,Graham2016} and the choice of general relativistic gauge~\cite{badurina2024signatureslinearizedgravityatom,Dimopoulos:2008hx}.
    %
}%
\label{fig:clocks-GW-concept}
\end{figure}

Atom interferometers, which employ lasers to split and recombine the wavefunctions of atoms, have optimal sensitivities to GWs with frequencies ${\cal O}(1)$~Hz~\cite{Badurina2020,Abe2021} and hence are well suited to explore the frequency gap between terrestrial and space-borne laser interferometers as seen in \cref{fig:gravitational-waves}.
With the gradiometer configuration shown in \cref{fig:clocks-GW-concept}, a differential, single-photon, pair of atom interferometers separated by a baseline $\sim$ \qty{1}{km} could achieve sufficient sensitivity to detect GWs~\cite{yu_gravitational_2011,graham_new_2013} with frequencies $\sim$  \qty{1}{Hz} that currently cannot be measured. Such detectors are also sensitive to theorized interactions between atomic constituents and bosonic dark matter fields with masses $\sim \qty{e-15}{eV}$~\cite{Arvanitaki:2016fyj}, with potential resolution significantly beyond existing experiments~\cite{Badurina2020}.

Long-baseline atom interferometers are under development by the AION (Atom Interferometer Observatory and Network)~\cite{Badurina2020} and MAGIS (Matter-wave Atomic Gradiometer Interferometric Sensor)~\cite{Abe2021} collaborations, and other projects worldwide~\cite{TVLBAIMOU}. These join other proposed approaches in the mid-frequency band, including space-based laser interferometers such as DECIGO~\cite{Kawamura:2011zz} and magnetically levitated superconducting test masses~\cite{paik_low-frequency_2016}, see Ref.~\cite{Sedda:2019uro} for a review. Realising the potential of atom-interferometer experiments will require overcoming many technical obstacles to reach the target sensitivity. One open question for these projects is whether the laser phase noise, which introduces noise on each individual atom interferometer that is orders of magnitude higher than the Standard Quantum Limit (SQL --- see Methods), will cancel sufficiently in the gradiometer configuration to reach the SQL.\@
While the gradiometer principle has previously been demonstrated in experiments using \ch{^{88}Sr}~\cite{hu_atom_2017}---or \ch{^{87}Rb}~\cite{snadden_measurement_1998,stray_quantum_2022}, to within known limitations~\cite{le_gouet_influence_2007}---in this work we quantify the extent of noise cancellation afforded by the scheme. We do this for the first time with the more demanding fermionic isotope \ch{^{87}Sr}, whose hyperfine structure and mHz-linewidth clock transition significantly complicate laser cooling and atom interferometry~\cite{mukaiyama_recoil-limited_2003,boyd_nuclear_2007,desalvo_degenerate_2010,akatsuka_optically_2017,hu_sr_2019}.
Despite these complications, \ch{^{87}Sr} is a natural choice for gravitational-wave detection, thanks to its near-ideal properties as an atomic clock isotope~\cite{Aeppli:24} and \qty{150}{\second} excited-state lifetime~\cite{lu_determining_2024}. These qualities are not shared by other candidate species such as $^{87}$Rb or $^{88}$Sr but are essential for very-long-baseline experiments, enabling the extension to space-scale baselines as proposed by the AEDGE (Atomic Experiment for Dark Matter and Gravity Exploration) project~\cite{AEDGE:2019nxb}. The same differential measurement configuration that enables gravitational-wave detection with \ch{^{87}Sr} also provides sensitivity to ultralight dark matter, which would induce coherent oscillations in the clock transition frequency across the detector baseline~\cite{Arvanitaki:2016fyj}.

We describe in this paper how the AION project has, for the first time, tested a gradiometer configuration in the laboratory using $^{87}$Sr. We combine atomic clock technology with atom interferometry, forming two macroscopically separated interferometers interrogated by a common clock laser. We find that our prototype detector reaches the SQL even in the presence of several radians of synthetic laser phase noise, emulating the conditions of a full-scale detector. Our results imply laser noise cancellation consistent with full common-mode rejection to within the measurement resolution of our experiment. Finally, we show that the same differential configuration allows the recovery of coherent time-dependent signals, even under conditions where a single interferometer would retain no recoverable phase information. While further work will be essential to demonstrate laser phase noise cancellation with larger numbers of atoms (for which the SQL is lower), and at longer baselines where wavefront propagation effects become relevant, our work verifies the principles underpinning long-baseline, single-photon atom interferometry, passing an important milestone on the road towards measurement of gravitational waves.

Analogously to the interference of light in a laser interferometer such as that used in the LIGO, Virgo and KAGRA experiments, atom interferometry relies on the interference of quantum matter waves. In the search for gravitational waves, both techniques probe a long baseline whose length in the proper detector frame is modulated by a gravitational wave, converting the variations in time-of-flight of light along this baseline to a variation of phase in an interference measurement---see \cref{fig:clocks-GW-concept}. For a discussion in a fully relativistic framework, see Refs.~\cite{Dimopoulos:2008hx,badurina2024signatureslinearizedgravityatom}. In laser interferometers, the interference is between light beams that travel along different paths. In atom interferometers, the interference is in the wavefunctions of atoms that are manipulated by laser pulses to follow spatially-separated paths before recombination.

\begin{figure}[t]
	\centering
    \includegraphics[width=\textwidth]{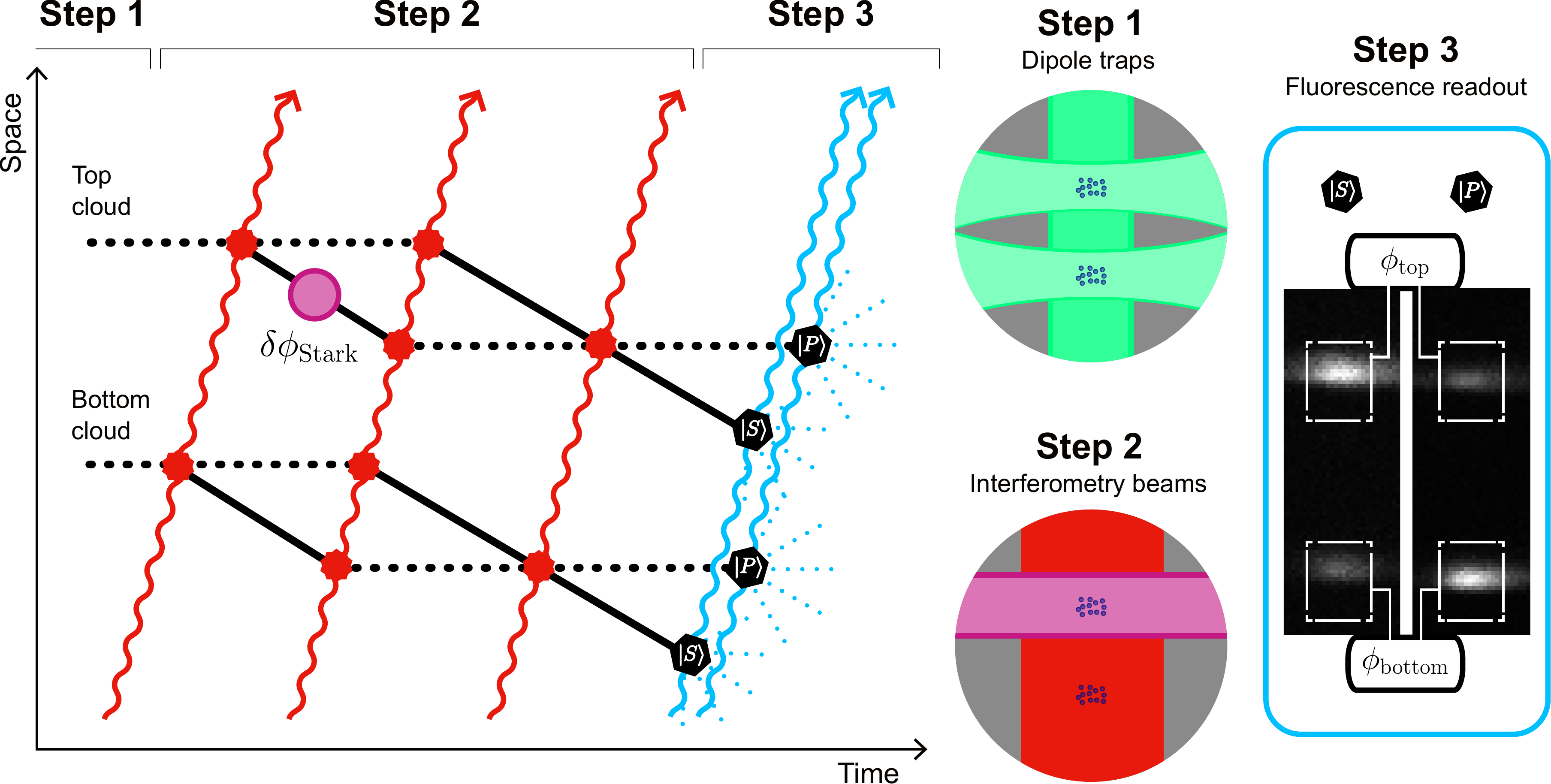}
\caption{%
    \textbf{Overview of steps in the experiment.} %
    \textit{(Main figure, left)} Spacetime diagram of the paths taken by the arms of each interferometer. Clock pulses (red) create a superposition of $^1$S$_0$ (solid) and $^3$P$_0$ (dotted) atomic states following both paths. For simplicity, we do not show the free-fall trajectories of the atoms. %
    \textit{(Step 1)} Two clouds of $^{87}$Sr atoms are confined, cooled and then released from crossed optical traps (green). %
    \textit{(Step 2)} An ultrastable clock laser (red) interrogates both clouds. An additional beam (purple) applies a phase shift $\dStark$ to just the top cloud, inducing a controllable differential phase between the two atom interferometers. 
    \textit{(Step 3)} The interferometer phases are determined by measuring the populations of the ground and excited states using fluorescence measurements. The fluorescence images shown here had an excitation fraction of \excitationFractionTop{} in the top cloud and \excitationFractionBottom{} in the bottom. See Methods for details of all these steps.%
}%
\label{fig:experimental-overview}
\end{figure}

In a single-photon atom interferometer, the atomic wavefunction is manipulated using pulses of light that drive a single-photon transition in the atom, often referred to as a clock transition. For the pulse sequence shown in \cref{fig:experimental-overview}, the phase of a single interferometer can be written in the simplified form
\begin{equation}
\phi =
\int_{-\infty}^\infty \omega_0\, g(t) \mathrm{d}t +
\phi_\text{laser} +
\phi_\text{other},
\end{equation}
where $\omega_0$  is the angular frequency of the atomic clock transition, $\phi_\text{laser}$ represents the total phase imprinted on the atoms due to the laser's phase during pulses, and $\phi_\text{other}$ comes from various sources such as static background gravitational or electromagnetic fields~\cite{Dimopoulos:2008hx,Abe2021,graham_new_2013,hogan_atomic_2011}, which do not play a role in the dark matter or gravitational wave detection.
$g(t)$ is determined by the relative states of the interferometer's upper and lower arms, where

\begin{equation}
g(t) = \begin{cases}
    -1 & \text{for $t$ between the first beamsplitter pulse and the mirror pulse,}\\
    +1 & \text{for $t$ between the mirror pulse and the final beamsplitter pulse,}\\
    0 & \text{otherwise.}
\end{cases}
\end{equation}

In long-baseline atom interferometry, a fundamental physics signal is extracted by taking the difference in phase $\dPhi = \phi_\text{top} - \phi_\text{bottom}$ between two atom interferometers separated by a long distance. To visualize the sensitivity of \dPhi{} to gravitational waves, the atom interferometers can be conceptualised as atomic clocks that are sensitive to small changes in the time taken for light to traverse the baseline~\cite{norcia_role_2017}.
The clocks ``tick'' while $g_\text{top}(t)$ and $g_\text{bottom}(t)$ are non-zero; these intervals are defined by the light pulse arrival times at each interferometer, so a modulation of the baseline $L$ by a gravitational wave alters the time counted by the clocks.
Alternatively, ultralight dark matter may cause small oscillations of atomic energy levels, affecting the tick-rate $\omega_0$ differently due to the time-delay between the two interferometers~\cite{bongs_high-order_2006,Graham:2015ifn,Arvanitaki:2016fyj,Badurina:2021lwr}. An important technical advantage of taking a \textit{differential} measurement is that the noise in the laser-induced phase $\phi_\mathrm{laser}$ cancels in common-mode: without laser noise cancellation, it would be unfeasible to achieve the ultimate target phase resolution of \qty{e-5}{\radian\per\sqrt{\hertz}} in the detector~\cite{Badurina2020} even using extremely low-noise lasers (see Methods).

Our tabletop prototype of a long-baseline atom interferometer detector is illustrated in \cref{fig:experimental-overview}. We operate a pair of crossed optical dipole traps, separated vertically by \qty{1}{\milli\meter}, containing clouds of fermionic $^{87}$Sr atoms at a temperature of approximately \qty{2}{\micro\kelvin}, loaded from a narrow-linewidth magneto-optical trap (see Methods for details). After the two clouds are released into free fall, an ultrastable clock laser (described in Ref.~\cite{stray_centralized_2024}) addresses the \ch{^1S0}~-~\ch{^3P0} optical clock transition. A first pulse (not shown) selects the slowest atoms from the falling clouds, and a sequence of three pulses then splits, reflects, and recombines the selected atoms to create two simultaneous Mach-Zehnder atom interferometers~\cite{kasevich_atomic_1991}. After the first beam-splitter pulse, we apply an additional, horizontal laser pulse, off-resonant from the \ch{^1S0}~-~\ch{^3P1} transition, to induce a controllable Stark shift \dStark{} to just one of the interferometers (see Methods). The same Stark-shifting pulse is applied in every shot of the experiment, biasing the phase offset between the interferometers to aid the data analysis.

\begin{figure}[t]
	\centering
	\includegraphics[width=\textwidth]{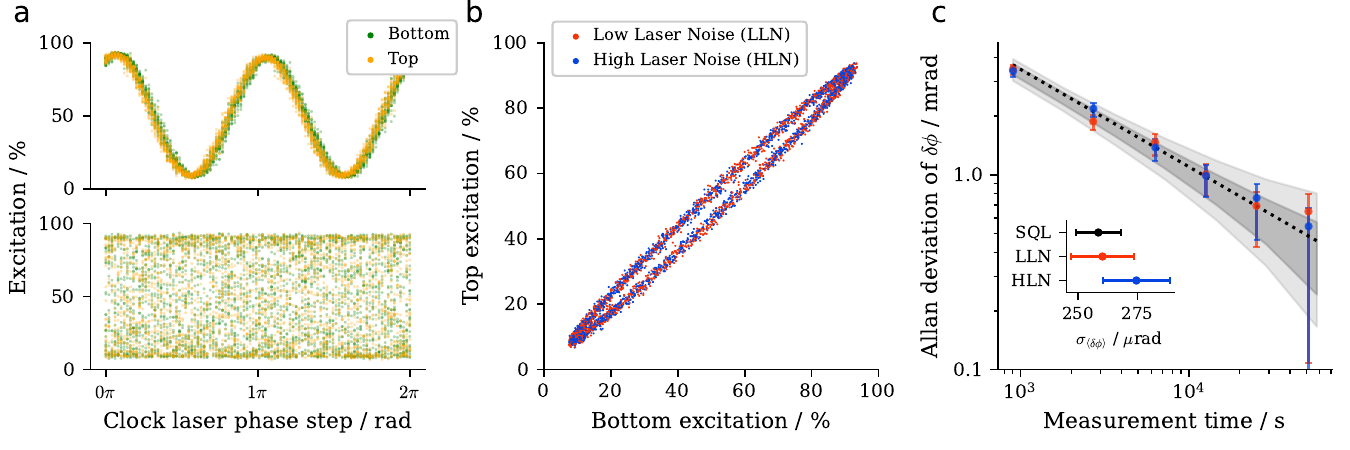}
    \includegraphics[width=\textwidth]{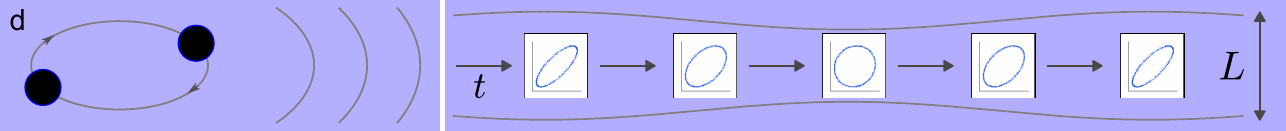}
    \caption{%
    \textbf{A test of laser noise rejection in the differential phase measurement.} %
    \textit{(a)}
    Clock atom interferometry fringes in the top and bottom atom clouds with a fixed Stark shift \dStark{} applied to the top cloud. The lower (upper) plot shows the fringes with (without) the addition of artificial laser noise (see Methods for details). %
    \textit{(b)} A correlation plot---or Lissajous figure---of the top and bottom atom interferometer signals, with (blue) and without (red) added laser noise. 
    \textit{(c)} Overlapping Allan deviations of the differential-phase time series $\dPhi(t_i)$, calculated using
    unbinned maximum likelihood estimation. The grey bands represent the 1-$\sigma$ and 2-$\sigma$ bounds on a SQL prediction from \numMCSims{} Monte Carlo simulations using independently measured experimental inputs, rather than a fit to the Allan deviation data. The black dotted line shows the expected averaging behaviour of white noise at the level defined by the theoretical SQL. %
    \textit{(c inset)} The standard errors $\sigma_{\left<\delta\phi\right>}$ of the differential phase measurements for the Low Laser Noise (LLN) and High Laser Noise (HLN) datasets, compared with the Cramer-Rao theoretical bound (see Methods).\@
    \textit{(d)} An illustration of how a passing low-frequency gravitational wave would modify the shape of the ellipse in the Lissajous figure: a gravitational wave would modulate \dPhi{} as the second derivative of the strain~\cite{graham_new_2013}.
    }
\label{fig:phase_resolution}
\end{figure}

To gather the datasets presented in \cref{fig:phase_resolution}, we scan the relative phases of the three clock pulses applied to both atom interferometers. \Cref{fig:phase_resolution}a shows the typical interference fringes we obtain. Using a $\pi$-pulse duration of \piPulseDuration{} and a free-fall time of $T=\sequenceTime{}$ between pulses, interferometer contrasts of \useVal{qpn-contrast-top} and \useVal{qpn-contrast-bottom} are observed. In order to simulate the effect of laser phase noise on a long-baseline atom interferometer, for one of the datasets we inject randomised phase steps to the clock laser between pulses of the atom interferometer sequence. This simulates the effect of laser phase error accumulated during the sequence, although it neglects its effect on the fidelity of mirror pulses. This is a reasonable representation of a long-baseline detector, since laser noise will be integrated over drop times of many seconds~\cite{Badurina2020}, amplifying its impact relative to our short sequence of \sequenceTime{} (see Methods for a calculation). The resulting individual interference fringes are shown in the lower panel of \cref{fig:phase_resolution}a, and are completely obscured by the injected noise. However, the differential phase \dPhi{} of the two interferometers can still be recovered using a maximum likelihood analysis~\cite{PDG} applied to the correlated excitation fractions from both interferometers. The likelihood model treats the shot-to-shot common phase as a nuisance parameter, which is marginalised (see Methods). 
The Lissajous correlation plot (\cref{fig:phase_resolution}b) provides a visualisation of the common-mode correlation.

To measure the impact of laser noise on the stability of the differential phase measurement, we compare measurements with the same applied differential phase \dStark{}, but with different levels of applied laser noise. We gather a ``Low Laser Noise (LLN)'' dataset in which only the intrinsic noise of the ultrastable clock laser is present~\cite{stray_centralized_2024} and a ``High Laser Noise (HLN)'' dataset with several radians of laser phase noise artificially added to each shot. The LLN and HLN data are interleaved shot-by-shot, with a total of \dataNumShots{} shots taken over a period of \dataDuration{}. For each of the two datasets, we extract a time-series of differential phases $\dPhi(t_i)$ using an unbinned maximum-likelihood analysis that operates on
blocks of \shotsPerBin{} excitation measurements from both  --- full details are available in Ref.~\cite{mle_upcoming}. In contrast to geometric ellipse fitting of binned Lissajous figures~\cite{Foster_2002}, the maximum likelihood analysis exhibits negligible bias and reduced statistical error~\cite{mle_upcoming}, remaining robust even in the fully phase-randomised regime.
\cref{fig:phase_resolution}c shows the Allan deviation~\cite{allan_statistics_1966} of these datasets; despite laser noise that completely obscures individual interferometer fringes, we observe no statistically significant increase in differential phase noise $\sigma_{\delta\phi}$ beyond the Cramer-Rao SQL~\cite{corgier_optimized_2025,pezze_quantum_2018} of $\sigma_{\dPhi} = \theoreticalSQLSingleShot{}$ per shot, determined by the \useVal{median-atoms-top} and \useVal{median-atoms-bottom} atoms measured in the top and bottom traps (see Methods). Extrapolating to the full experimental run of \dataNumShots{} shots, split between the HLN and LLN datasets, we project a SQL of $\sigma_{\left<\delta\phi\right>}=\theoreticalSQLWholeDataset$ in the average differential phase $\left<\delta\phi\right>$ for either dataset.

To quantify this rejection, we use the maximum likelihood analysis to infer the noise levels of the measured \dPhi{} time-series in both cases. Extrapolating to the whole datasets, we determine that the standard deviations $\sigma_{\left<\delta\phi\right>}$ of \dPhi{} are consistent with the SQL in both the LLN and the HLN dataset, with
$\sigma_{\left<\delta\phi_\text{LLN}\right>}-\sigma_{\left<\delta\phi_\text{SQL}\right>} =
\lowLaserNoiseSigmaMinusSQL$
and
$\sigma_{\left<\delta\phi_\text{HLN}\right>}-\sigma_{\left<\delta\phi_\text{SQL}\right>} =
\highLaserNoiseSigmaMinusSQL$
(\cref{fig:phase_resolution}c, inset).
Crucially, we observe no statistically significant increase in noise despite the addition of several radians of shot-to-shot laser phase noise in the HLN dataset, with
$\sigma_{\left<\delta\phi_\text{HLN}\right>} - \sigma_{\left<\delta\phi_\text{LLN}\right>} = \differenceSigmaWholeDataset$,
consistent with zero additional differential-phase noise within uncertainty, despite completely scrambled interferometer phases.

In addition to the Cramer-Rao SQL, we validate the statistical performance of the estimator (bias and coverage) at the SQL using Monte Carlo simulations matched to experimental conditions (see Methods). The Monte Carlo band shown in Fig.~\ref{fig:phase_resolution}c is not a fit to the Allan deviation data: it is a prediction constructed from independently measured atom-number statistics and interferometer contrasts, processed through the same phase-extraction pipeline as the real data. Its agreement with the measured Allan deviation therefore constitutes a non-trivial closure test of the statistical model and is consistent with SQL-limited operation.

\begin{figure}[tb]
	\centering
    \includegraphics[width=\textwidth]{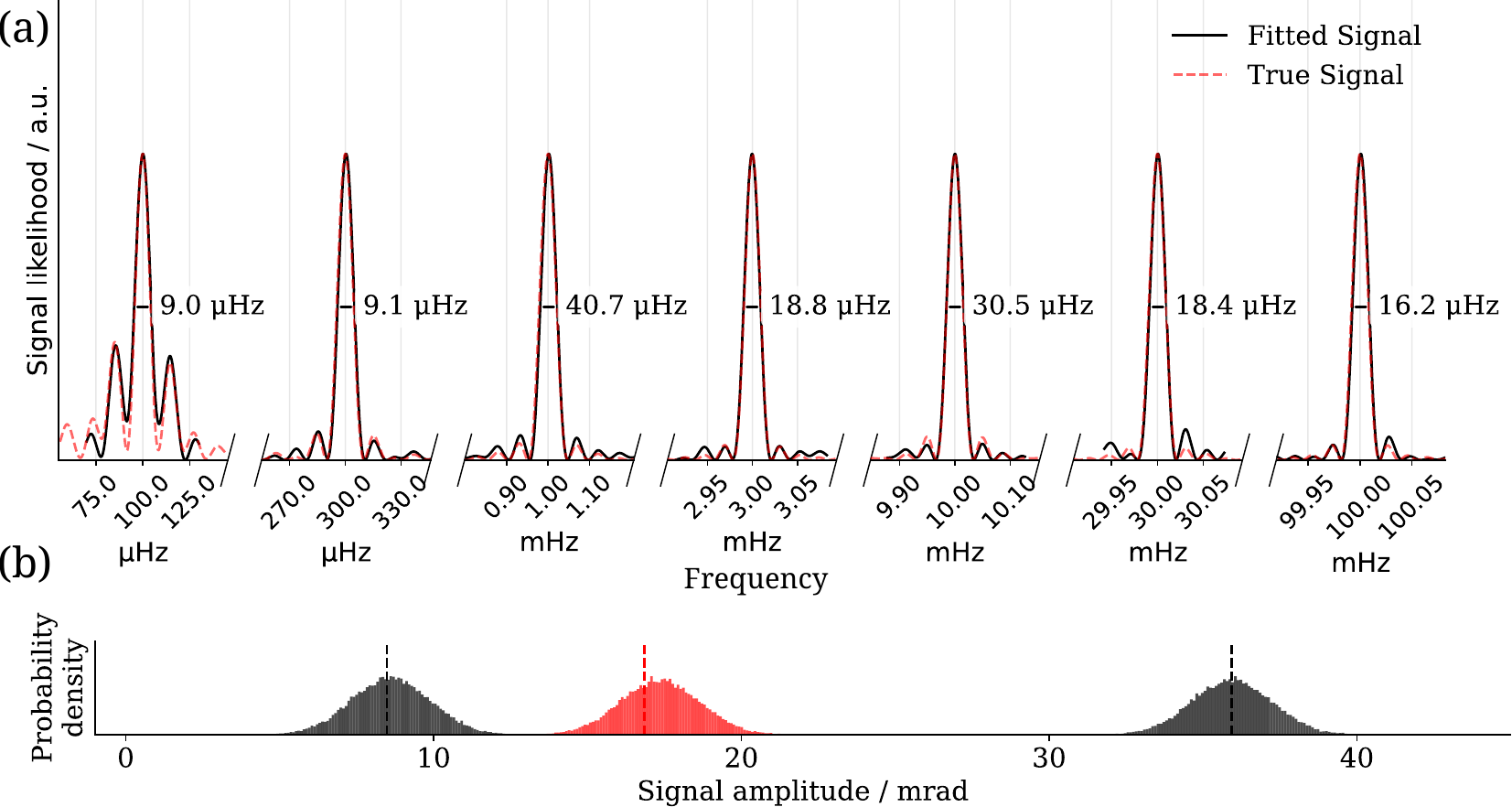}
\caption{%
    \textbf{Signal recovery in the presence of high laser noise.} %
    An oscillating signal is injected into the differential interferometer to emulate signals from fundamental sources. We extract maximum-likelihood estimates for the signal's frequency $\omega$, amplitude $A$, and phase $\chi$ using an unbinned fit over the whole dataset, with all datasets using fully randomized laser phase noise as described previously to match conditions in a true long-baseline interferometer.
    \textit{(a)} Signal recovery over 7 geometrically-spaced frequencies, covering our prototype detector's Nyquist band. Each injected signal has a nominal amplitude of \injectedSignalAmplitudeForFreqs{}. Solid lines show the likelihood of a signal's presence extracted from the interferometer's output. Dotted lines show
    a periodogram of the true injected signal (artifacts here are due to our finite-length, non-deterministically-spaced samples). Despite a dynamic range spanning several orders of magnitude, our detector can resolve injected signals with a $\mathcal{Q}$-factor of over \num{5000}, limited by integration time.
    \textit{(b)} Inferred signal amplitudes for three sample signals.
    The histograms show Monte-Carlo-simulated probability distributions for SQL-limited measurements for the three cases, seeded with their expected amplitudes.
    Fit results from the interferometer datasets are marked with dotted lines, in agreement with our simulated SQL-limited uncertainties. For this test, we used injected signal frequencies of either
    \qty{1}{\milli\hertz} (black) or \qty{100}{\milli\hertz} (red)
    to cover the detector's range.
    }%
\label{fig:signal-extraction}
\end{figure}

Beyond the extraction of a constant differential phase used to quantify laser-noise cancellation, the same maximum likelihood framework used in an unbinned mode, but with time-dependent \dPhi{}, enables hypothesis testing for oscillatory signals in the differential measurement~\cite{mle_upcoming}. This provides a proof-of-principle that physically relevant signals---such as those expected from gravitational waves or ultralight dark matter---can be extracted in a differential atom-interferometer configuration under conditions where signal recovery would be impossible using a single interferometer alone. Crucially, this signal-fitting approach remains effective in the shot-to-shot phase-randomised (high laser-noise) regime: a single atom interferometer contains no recoverable phase information in this regime, whereas the differential measurement retains statistically recoverable sensitivity to coherent signals through common-mode noise rejection.

In \cref{fig:signal-extraction}, we test this directly with controlled signal injection and recovery under fully phase-randomised conditions. We inject controlled sinusoidal phase modulations via the off-resonant Stark shift \dStark{} applied to the top interferometer, and we analyse the resulting excitation fraction record with the same unbinned likelihood model but applied over the whole dataset. Defining $\delta\phi(t) = \delta\phi_0 + A \sin(\omega t + \chi)$, we demonstrate signal recovery at representative test frequencies spanning the range \qtyrange{e-4}{e-1}{\hertz}, with results shown in \cref{fig:signal-extraction}. These are compared with the signal recovery of a perfect, noiseless detector limited only by integration time, with excellent agreement.
These frequencies lie within the measurement bandwidth of the present prototype, set by the shot cycle time $\sim$ \qty{3}{\second} and the total run duration (hours to $\sim1$~day). This choice reflects the prototype operating conditions rather than a fundamental limitation of the method; in a future long-baseline detector, the sensitive band is shifted by design through the interrogation time, baseline, and repetition rate into the mid-frequency regime.

We also test amplitude recovery by probing signals at fixed frequency of either
\qty{1}{\milli\hertz} or \qty{100}{\milli\hertz},
with varying amplitudes (\cref{fig:signal-extraction}b). We verify that the recovered modulation amplitude scales linearly with the applied Stark-shift duration and use this nominal calibration to seed \num{e6} Monte-Carlo simulations for each scenario to understand our detector's sensitivity, assuming no noise other than the SQL. All signals are recovered with SQL-limited resolution
and correctly favour the null hypothesis when no signal is present.

All signal-injection experiments were performed in the high laser-noise regime, emulating the behaviour of a long-baseline detector. We find that the differential channel yields statistically significant recovery of injected signals, fully rejecting laser phase noise in the signal recovery.


The successful integration of clock transition techniques with atom interferometry is an important milestone on the path towards their joint implementation in quantum sensors with applications in fundamental physics. These include not only the detection of ultralight dark matter and gravitational waves~\cite{Badurina2020,Abe2021,Badurina:2021rgt}, but also tests of equivalence principles~\cite{asenbaum_atom-interferometric_2020,tarallo_test_2014} and measurements of the fine structure constant~\cite{schelfhout_single-photon_2024}. The construction of long-baseline detectors will also spur advanced quantum sensing with applications in navigation, geodesy, and resource exploration (see, e.g., Ref.~\cite{stray_quantum_2022}).

There are many further technical hurdles to be overcome before a long-baseline detector can be realised. These include the development of more intense sources of cold atoms, the extension to longer baselines while controlling associated systematic shifts~\cite{Abe2021,hogan_atomic_2011}, large momentum transfer from the laser to the atoms~\cite{rudolph_large_2020} and the use of squeezed atomic states~\cite{corgier_optimized_2025}. All of these are the subjects of R\&D programmes in several groups within the international Terrestrial Very-Long-Baseline Atom Interferometry Proto-Collaboration~\cite{TVLBAIMOU}. Nevertheless, the experimental techniques already demonstrated here open up exciting new avenues for scientific exploration that range from probing the fundamental laws that govern our Universe to enhancing quantum sensors. 

\FloatBarrier{}

\bibliography{bibliography}

\section*{Methods}

\textbf{Cooling sequence:} The cold atom apparatus used in this experiment has previously been described in Refs.\ \cite{pasatembou_progress_2024, stray_centralized_2024}. To prepare samples of cold $^{87}$Sr, the atoms are first collected for \qty{1.5}{s} into a blue 3D magneto-optical trap (MOT) that uses the \ch{^1S0}~-~\ch{^1P1} transition at \qty{461}{nm} and a field gradient of \qty{3.5}{\milli\tesla\per\centi\meter}. Atoms that leak into the metastable $^3$P$_2$ manifold are recycled into the MOT using repump lasers at \qty{679}{\nano\meter} and \qty{707}{\nano\meter}. For efficient repumping of $^{87}$Sr, frequency sidebands at \qty{585}{MHz} and \qty{487}{MHz} are applied to the \qty{707}{\nano\meter} light using an electro-optic modulator, creating frequency components near-resonant with transitions from all five hyperfine manifolds of $^3$P$_2$~\cite{yang_high_2015}.

After the blue MOT is switched off, the atoms are captured into a red MOT operating on the $\ch{^1S0}\, F = 9/2$ to $\ch{^3P1}\, F' = 11/2$ transition at \qty{689}{\nano\meter}, using a field gradient of \qty{390}{\micro\tesla\per\centi\meter}. Sidebands at \qty{1463.265}{\mega\hertz} are applied to the \qty{689}{\nano\meter} light using a resonant electro-optic modulator, addressing the $F = 9/2$ to $F' = 9/2$ transition in order to stir the atoms between Zeeman sublevels of the ground state, mitigating losses into sublevels where atoms are weakly confined~\cite{mukaiyama_recoil-limited_2003}. During the first \qty{220}{\milli\second} in the red MOT an intensity of $1800I_\mathrm{sat}$ is used in each of the six MOT beams, where $I_\mathrm{sat}=\qty{3}{\micro\watt\per\square\centi\meter}$ is the saturation intensity of the \qty{689}{\nano\meter} transition. To capture the wide range of Doppler-shifted atoms released from the blue MOT, sawtooth-wave modulation is applied to the \qty{689}{\nano\meter} light at a sweep frequency of \qty{20}{\kilo\hertz} and a peak-to-peak sweep range of \qty{6}{\mega\hertz}~\cite{snigirev_fast_2019}. For the following \qty{100}{\milli\second}, while in the ``narrowband'' red MOT, the sawtooth frequency modulation is switched off and the intensities of the six MOT beams are ramped linearly from $490I_\mathrm{sat}$ to $40I_\mathrm{sat}$. In order to help support the atoms against the force of gravity, a seventh, unbalanced MOT beam---the ``up'' beam---is introduced in the vertical direction during the narrowband MOT.\@ The up beam is necessary for creating narrowband red MOTs below $100I_\mathrm{sat}$ without significant atom loss. Upon completion of the narrowband red MOT, the atoms have a temperature of \qty{2}{\micro\kelvin} and are compressed into a region comparable in size to the optical dipole trap.

\textbf{Dipole trap and state preparation:} Two crossed optical dipole traps, separated vertically by \qty{1}{mm}, are formed by separate \qty{2.5}{W} horizontal beams at \qty{1064}{nm} with horizontal and vertical $1/e^2$ radii of \qty{220}{\micro\meter} and \qty{23}{\micro\meter} respectively, crossed with a shared \qty{840}{mW} vertical beam at \qty{813}{nm} with $1/e^2$ radii of \qty{60}{\micro\meter} in both transverse axes. Overlapped with the top crossed dipole trap, a \qty{4}{\milli\watt} transparency beam at \qty{488}{\nano\meter}, detuned by \qty{25}{\giga\hertz} from the $5s5p\,\ch{^3P1}-5s5d\,\ch{^3D2}$ transition, is applied with a $1/e^2$ radius of \qty{40}{\micro\meter} to protect the atoms from scattering \qty{689}{\nano\meter} light after they are loaded into the top crossed dipole trap region.

Immediately after the free-space red MOT stages described above, the dipole trapping beams, the transparency beam, and repumpers at \qty{679}{\nano\meter} and \qty{707}{\nano\meter} are switched on; the red MOT is then held for \qty{100}{\milli\second} in a ``top-trap loading'' stage, during which the bias magnetic fields, beam intensities, and detunings of the red MOT are optimised to load the atoms into the upper of the two dipole traps. During the top-trap loading stage, the red MOT intensity is linearly ramped from $20I_\mathrm{sat}$ to $4I_\mathrm{sat}$ to steadily reduce the atom temperature. Next, to load the bottom optical dipole trap, the red MOT is released for \qty{3}{\milli\second} by switching off the \qty{689}{nm} beams. During this time, the cold atoms already in the top trap are held in place, while the hotter atoms fall towards the bottom trap. While the atoms fall, the vertical bias magnetic field is stepped such that the zero of the quadrupole magnetic field is close to the bottom dipole trap. After \qty{3}{\milli\second} of free fall, the red MOT beams are switched back on for \qty{100}{\milli\second} in a ``bottom-trap loading'' stage using the same parameters as the top-trap loading stage, except for the different bias magnetic field. All but the hottest atoms in the top trap remain in the top trap during the bottom-trap loading stage, since they are protected by the \qty{488}{\nano\meter} transparency beam against scattering \qty{689}{\nano\meter} light.

After both dipole traps are loaded the MOT beams are switched off, a horizontal bias field is applied, and the trapped atoms are optically pumped into the stretched state $M_F = 9/2$ by applying a horizontal bias field of \qty{38}{\micro\tesla} and delivering a \qty{20}{ms} pulse of circularly-polarised light at \qty{689}{nm}, resonant with the $^1$S$_0$ $F = 9/2$ to $^3$P$_1$ $F' = 9/2$ transition. During the optical pumping, sawtooth-wave frequency modulation is applied to the \qty{689}{nm} light at a rate of \qty{30}{\kilo\hertz} over a range of \qty{6}{\mega\hertz}. Finally, all beams except the dipole trap are switched off, and the bias magnetic field is adiabatically ramped to the final field used for atom interferometry: \qty{31}{\micro\tesla} aligned with the linear polarisation of the vertical \qty{698}{nm} clock beam.

\textbf{Velocity selection on the clock transition:} The clock beam at \qty{698}{nm} propagates vertically upward through both dipole trap regions with a waist of \qty{600}{\micro\meter}. The clock laser linewidth is verified against an independent cavity-stabilised laser as below \qty{2}{\hertz}, prior to delivery of the light to atoms through an uncompensated \qty{10}{m} fibre~\cite{stray_centralized_2024}. Clock spectroscopy sequences are carried out immediately after atoms are released from both dipole traps, and then the excitation fraction is detected using a \qty{200}{\micro\second} fluorescence pulse at \qty{461}{nm} to detect the number of atoms in the ground state $^1$S$_0$, followed by a \qty{3.5}{\milli\second} repumping pulse at \qty{679}{nm} and \qty{707}{nm} and another \qty{200}{\micro\second} fluorescence pulse at \qty{461}{nm} to detect atoms that are in the $^3$P$_0$ state after the interferometer sequence. Scattered light from each \qty{461}{nm} spectroscopy pulse is gathered in separate exposures on an \textit{Andor iXon Ultra 897} EMCCD camera, and a separate EMCCD image without atoms present is used to subtract background counts.

At the maximum available clock power of \qty{640}{\milli\watt}, a Rabi $\pi$-pulse time of \qty{44}{\micro\second} is measured. However, the clock transition is observed to have a peak excitation fraction of 0.3, and a Doppler-broadened linewidth of \qty{60}{\kilo\hertz}---considerably larger than the \qty{20}{\kilo\hertz} Fourier limit. In order to improve the fidelity of the Rabi pulses in the atom-interferometer sequence, a velocity selection procedure is used. The clock beam is pulsed on for \qty{200}{\micro\second} at \qty{20}{\milli\watt}, implementing a $\pi$ pulse that excites the slowest atoms to the upper clock state $^3$P$_0$. The atoms in the ground state are then pushed away using a \qty{500}{\micro\second} pulse at \qty{461}{nm}, leaving only the slow atoms in the \ch{^3P0} state to enter the interferometer sequence. After this velocity selection sequence, a resonant, \qty{44}{\micro\second} Rabi $\pi$-pulse yields a peak de-excitation fraction of 90\%.

\textbf{Clock atom interferometry:} The clock atom interferometry consists of a sequence of three resonant pulses on the \qty{698}{\nano\meter} clock transition, with pulse areas $\pi/2-\pi-\pi/2$, a $\pi$ pulse time of $ t_\pi = \piPulseDuration{}$, and a dark time of $T = \sequenceTime{}$ between each consecutive pulse. For the data in \cref{fig:phase_resolution}, the phase of the clock light is always stepped deterministically during the dark times such that the phase of the first, second and third pulses are $0$, $\phi$ and $4\phi$ respectively, with $\phi$ ranging from $0$ to $2\pi$ in \useVal{num-phase-steps} steps in a randomized order.
Each datapoint in the right-hand side of \cref{fig:phase_resolution} results from $2\times\useVal{num-phase-steps}$ samples, interleaved between ``High Laser Noise'' and ``Low Laser Noise'' samples.
``High Laser Noise'' samples have additional phase steps applied during the interferometer dark times (see \cref{fig:experimental-overview}), drawn independently from a Gaussian distribution with a standard deviation of
\useVal{stddev-random-phase} and mean of $\qty{0}\ {\rm rad}$. It is important to distinguish between two types of randomisation employed in this work. In both LLN and HLN datasets, the clock laser phase is scanned deterministically through 100 values in randomised order; this scan-order randomisation ensures that any spurious time-oscillatory signals, such as 50 Hz from room lights, will not be aliased to look like apparent fringes. In the HLN dataset, we additionally apply large, uncorrelated phase jumps between shots, which fully randomise the absolute phase of each individual interferometer on a shot-by-shot basis. This per-shot phase randomisation mimics the regime expected in long-baseline atom interferometers, where integrated laser frequency noise over multi-second interrogation times will produce phase excursions of many radians (see below). Under these conditions, a single atom interferometer retains no recoverable phase information, so this provides a stringent test of the differential measurement's noise rejection capability. The phase randomisation fully masks the fringes in each individual interferometer, but does not affect the measurement of the relative phase of the two interferometers.

\textbf{Laser phase noise estimate for a km-scale detector:} The phase noise imparted onto the atoms by the laser can generally be calculated from the spectral density of frequency fluctuations on the laser~\cite{le_gouet_limits_2008}. In our prototype, the laser phase imprinted on each atom interferometer in one repetition of the interferometer sequence beginning at time $t$ is approximately $\phi_\mathrm{laser} = \varphi(t) - 2\varphi(t+T) + \varphi(t+2T)$, where $\varphi(t)$ is the time-dependent phase of the laser field oscillating as $\cos{(kz-\omega_0 t + \varphi(t))}$, and where the approximation holds in the limit of short beamsplitter and mirror pulses separated by a dark time $T$~\cite{Dimopoulos:2008hx}. Treating $\varphi(t)$
as a stationary noise process with one-sided power spectral density $S_\varphi(f)$, and applying the optical Wiener-Khinchin theorem~\cite{steck_quantum_2024}, we observe a variance in interferometer laser phase:

\begin{align*}
\left<\phi^2_\mathrm{laser}\right> &= \left<\left( \varphi(t) - 2\varphi(t+T) + \varphi(t+2T) \right)^2\right> \\
&= 6\left<\varphi(t)\varphi(t)\right> - 8\left<\varphi(t)\varphi(t+T)\right> +  2\left<\varphi(t)\varphi(t+2T)\right> \\
&= \int_0^\infty S_\varphi(f)\left[6 - 8\cos(2\pi f T) + 2\cos(4\pi f T) \right]df
\end{align*}

For a future long-baseline atom interferometer, we consider a model for the clock laser as a thermal-noise-limited, cavity-stabilised laser~\cite{numata_thermal-noise_2004}, with a flicker frequency noise spectrum of the form $S_\varphi(f) = S_\varphi(f=\qty{1}{\hertz})\times\left(\qty{1}{\hertz}/f\right)^3$. Propagating this functional form through the above equation, the standard deviation of the interferometer laser phase simplifies as $\sqrt{\left<\phi^2_\mathrm{laser}\right>} = 4\pi T \sqrt{ \ln(2)}\sqrt{S_\varphi(\qty{1}{\hertz})}$. To provide an optimistic numerical estimate of the laser phase, we assume a laser noise spectrum at the limit of current laser technology, with fractional frequency noise $S_y(f) = ({10^{-33}/f})$/Hz~\cite{oelker_demonstration_2019}. For the $^{87}$Sr clock transition at \qty{698}{\nano\meter}, the corresponding noise spectral density of clock laser phase fluctuations would be $\sqrt{S_\varphi(\qty{1}{\hertz})} = \qty{14}{\milli\radian\per\sqrt{\hertz}}$, resulting in a standard deviation in interferometer laser phase $\sqrt{\left<\phi^2_\mathrm{laser}\right>} = \qty{710}{\milli\radian}$ for $T = \qty{5}{\second}$, the interferometer time projected for a km-scale detector~\cite{Badurina2020}. Even for an interferometer repetition rate of several shots per second, the laser phase noise imprinted on each individual interferometer is therefore far above the level needed to reach the ultimate target phase resolution of \qty{e-5}{\radian\per\sqrt{\hertz}}~\cite{Badurina2020}, highlighting the need for laser noise cancellation in the differential phase \dPhi{}.

Compounding the requirements for laser phase noise cancellation, a large momentum transfer of $n\sim10^4$ photon recoils is targeted for long-baseline detectors~\cite{Badurina2020}, enhancing the detector sensitivity but imprinting laser phase noise $n$ times onto each atom interferometer~\cite{Dimopoulos:2008hx}. Taking into account the large momentum transfer, long-baseline interferometers are likely to be in the fully phase-randomised regime explored by the ``High Laser Noise'' dataset in this work.

\textbf{Differential bias phase:}
In order to induce a consistent relative phase offset between the top and bottom atom interferometers, an additional, horizontal \qty{689}{\nano\meter} Stark shifting pulse is applied to the top interferometer only, for \qty{30}{\micro\second} during the gap between the first $\pi/2$-pulse and the middle $\pi$-pulse. The Stark shifting beam is detuned by \useVal{stark-pulse-detuning} from the $^1$S$_0$ $F = 9/2$ to $^3$P$_1$ $F' = 11/2$ transition, with a waist of \qty{500}{\micro\meter} and a power of \qty{1}{\milli\watt}, inducing a phase shift specifically on atoms in the ground state (the lower arm) of the top interferometer. For the data in this paper, the Stark shifting pulse is used to generate a bias differential phase \dStark{} between the top and bottom interferometers, causing the data to lie on a Lissajous ellipse (see Fig.~\ref{fig:phase_resolution}b) rather than a straight line, thus containing more information about the differential phase \dPhi{}. A non-zero differential bias phase is required for efficient, low-error extraction of \dPhi{}, whether \dPhi{} is extracted using a maximum likelihood estimator or the ellipse-fit method. In a long-baseline detector, the dark matter or gravitational wave signal would induce fluctuations in the ellipse fit angle, on top of the static bias.

\textbf{Experimental control:}
Electronic control signals are produced through the FPGA-based experimental
control platform ARTIQ~\cite{jordens2023artiq}. Control software is written in
Python and is available open-source at Ref.~\cite{zenodo_repository_raw}.

\textbf{Phase extraction:}
Full details of this analysis are available in Ref.~\cite{mle_upcoming}.
We extract both constant differential phases (used to quantify laser-noise cancellation) and oscillatory signal components using a unified unbinned maximum likelihood analysis. For each experimental shot $i$, we model the measured excitation fractions $(y_{A,i}, y_{B,i})$ from the two interferometers $A$ and $B$ as noisy observations of sinusoidal interferometer responses that share a shot-dependent common phase $\phi_i$ but differ by a differential phase $\delta\phi(t_i)$. The common phase $\phi_i$ is treated as a nuisance parameter and marginalised, yielding a likelihood that depends only on the differential phase. In practice, we use this marginalised likelihood for inference: we report point estimates from maximisation of the marginal likelihood,
and compute uncertainties from repeated Monte Carlo simulations performed with matching parameters and analysed using the same analysis pipeline, following a hybrid Bayesian--frequentist approach commonly used in precision measurements and particle physics.

The per-shot likelihood is obtained by numerical integration over the common phase using a uniform prior on $[-\pi,\pi]$:
\begin{equation}
\mathcal{L}_i = \int_{-\pi}^{\pi}\frac{d\phi}{2\pi}\,\mathcal{N}(y_{A,i}|p_A(\phi),\sigma_{A,i}^2)\,\mathcal{N}(y_{B,i}|p_B(\phi,\delta\phi_i),\sigma_{B,i}^2),
\label{eq:likelihood}\end{equation}
where $\mathcal{N}(\cdot|\mu,\sigma^2)$ denotes a Gaussian probability density. The response functions $p_A$ and $p_B$ are sinusoidal fringe models of the form $p_A(\phi) = p_{0,A} + \frac{\mathcal{C_A}}{2}\cos\phi$ and $p_B(\phi) = p_{0,B} + \frac{\mathcal{C_B}}{2}\cos\left(\phi+\delta\phi\right)$, parameterised by offsets $p_{0,\{A,B\}}$ and contrasts $\mathcal{C}_{\{A,B\}}$, with noise variance $\sigma_{\{A,B\}}^2 = p_{\{A,B\}}(1-p_{\{A,B\}})/N_{\{A,B\}}$ describing the standard quantum limit resulting from the measured $N_{\{A,B\}}$ atoms in the two interferometers. This marginalisation enables robust inference even when individual interferometer fringes are fully washed out by laser phase noise.

\textit{Mode 1: differential phase stability analysis.} For the stability analysis (Allan deviation) presented in \cref{fig:phase_resolution}c, we estimate a piecewise-constant \dPhi{} over consecutive blocks of \shotsPerBin{} shots.

\textit{Mode 2: oscillatory signal analysis.} For the oscillatory-signal searches presented in \cref{fig:signal-extraction}, we parameterise the differential phase as $\delta\phi(t) = \delta\phi_0 + S\sin(\omega t) + C\cos(\omega t)$. This parameterisation captures the leading-order differential-phase response expected from both gravitational waves and ultralight dark-matter fields, which induce coherent oscillations via modulation of the effective light propagation time or atomic transition frequency. Signal significance is quantified using a likelihood-ratio test statistic comparing the best-fit signal model to the null hypothesis ($C = S = 0$). When scanning over frequency, we calibrate the null distribution of the test statistic with Monte Carlo simulations to account for the trials factor. In the absence of an injected signal, the framework correctly favours the null hypothesis, providing a statistically well-defined reference for future sensitivity studies. $C$ and $S$ can be converted to amplitude $A$ and phase $\chi$ using the formulae
\[
A = \sqrt{C^2 + S^2} \quad,
\\
\chi = \atantwo(-C/S)  \quad.
\]


The resolvable frequency band in the prototype is determined by the experiment's effective sampling interval (set by the average shot cycle time) and the observation duration. At low frequencies, sensitivity is limited by the finite run duration; at higher frequencies, it is limited by the shot rate and dead time. The injected-signal tests therefore probe the band where the prototype has statistical power over hour-to-day records. In a long-baseline detector, the same analysis framework applies but the effective response and optimal band are engineered via interrogation time, repetition rate, and baseline, shifting the instrument's peak sensitivity into the mid-frequency regime. Accordingly, the resolvable frequency band is instrument-dependent: the frequency band of the prototype implementation does not represent an intrinsic limitation of differential atom interferometry nor of the analysis framework itself.

\textbf{Data filtering:}
The \qty{461}{nm}, \qty{689}{nm}, and \qty{698}{nm} laser locks are monitored throughout the experiment. Experiment runs in which one or more of these locks fails, or in which the observed number of atoms in either trap is below a manually-set threshold near 60\% of the median number of atoms, are considered invalid and excluded from the data.

\textbf{Number of atoms:}
Atoms are detected at the end of atom-interferometer sequences through fluorescence imaging on an EMCCD camera. Under the assumption that fluorescence scales linearly with number of atoms, the fluorescence signal can be converted to number of atoms using a calibration derived from absorption imaging of clouds of atoms prepared under identical conditions as those used for the atom interferometry. The number of atoms $N$ in the calibration dataset is extracted from the raw absorption images through the relation $N \sigma(\omega) = \int OD(x,y) dxdy$~\cite{hueck_calibrating_2017}, where $OD(x,y)$ is the optical depth of the sample at transverse position $x,y$ in the absorption probe beam, and $\sigma(\omega)$ is the absorption cross-section of the $^{87}$Sr atoms at the laser frequency $\omega$.

Since the hyperfine shifts of the states $^1$P$_1$ $F=7/2$, $9/2$, and $11/2$ are respectively $+37$, $-23$, and $-6$~MHz~\cite{kluge_levelcrossing_1974}, which are significant compared with the \qty{30.5}{\mega\hertz} natural linewidth of the $^1$P$_1$ state~\cite{nagel_photoassociative_2005}, the absorption cross-section $\sigma(\omega)$ in $^{87}$Sr is generally polarisation- and $M_F$-dependent. To avoid reliance on direct measurements of the polarization of our absorption probe light and the $M_F$ state of the atoms, we instead measure the absorption amplitudes of the three lines from $^1$S$_0$ to $^1$P$_1$ $F=7/2$, $9/2$, and $11/2$ by carrying out spectroscopy over a $\pm$\qty{120}{\mega\hertz} range of detunings, using samples of atoms pumped into $M_F=9/2$ using the same preparation sequence as for the atom-number calibration and the atom-interferometry datasets. We fit the peak amplitudes $\sigma_{7/2}$, $\sigma_{9/2}$, and $\sigma_{11/2}$ of the three Lorentzians to the absorption spectroscopy data, using fixed literature values for the linewidths and the hyperfine splittings between the Lorentzians~\cite{kluge_levelcrossing_1974,nagel_photoassociative_2005}. Finally, we can calibrate the optical depth per unit atom using the identity that the sum of the peak absorption cross-sections must match the resonant absorption cross-section for the simpler isotopes with zero nuclear spin, i.e. $\sum_F\sigma_F = \sigma_0 = 3\lambda^2/2\pi$~\cite{hueck_calibrating_2017}. We obtain a total atom number uncertainty of
\qty{\useVal{atom-number-sigma-rel}}{\percent}
during the atom-interferometry datasets, dominated by the uncertainty in the difference in atom number between the calibration dataset and the fluorescence dataset.

During the HLN/LLN dataset, the median number of atoms in the top trap was \useVal{median-atoms-top} atoms and in the bottom \useVal{median-atoms-bottom} atoms. The number of atoms in each trap fluctuated during the \dataDuration{} dataset, with a maximum deviation of \useVal{peak-dev-percent} from the median. No significant difference in atom number was observed between shots with and without induced phase noise.

\textbf{Extraction of noise levels:}
To estimate any additional noise in our measurement of \dPhi{} caused by injecting laser noise, we apply the maximum likelihood phase-extraction method independently to both the LLN and HLN datasets. The time-series of extracted phases from \shotsPerBin{}-shot blocks is modelled as
\[
    \dPhi(t_i) \sim \mathcal{N}\!\left(\dPhi_0 ,\ \sigma_{\dPhi}^2 \right)
    \quad,
\]
where $\mathcal{N}(\mu, \sigma^2)$ denotes a normal distribution with mean $\mu$ and standard deviation $\sigma$. We use a No-U Turn Markov-chain Monte-Carlo method implemented using the PyMC package~\cite{Abril-Pla2023} to sample from the posterior distribution of $\sigma_{\dPhi}$, reporting the mean and \qty{68}{\percent} credible intervals as $\sigma_\text{LLN} = \lowLaserNoiseSigmaOneBlock$ and $\sigma_\text{HLN} = \highLaserNoiseSigmaOneBlock$.
To compare these to the standard deviations of Monte Carlo simulations with only SQL present (see below), we calculate from these per-block standard deviations the standard error on the mean over the whole dataset, giving
$\sigma_{\left<\delta\phi_\text{LLN}\right>}= \lowLaserNoiseSigmaWholeDataset$
and
$\sigma_{\left<\delta\phi_\text{HLN}\right>}= \highLaserNoiseSigmaWholeDataset$.

\textbf{Theoretical SQL:} We define the SQL as the Cramer-Rao bound to the per-shot phase noise $\sigma_{\delta\phi}$, calculated using the simple likelihood model in \cref{eq:likelihood} in which quantum projection noise is the only included noise process. The Cramer-Rao bound is a lower limit to $\sigma_{\delta\phi}$ for any unbiased estimator of \dPhi{}, regardless of the \dPhi{} extraction technique used, and is used as a rigorous benchmark in differential atom interferometers and quantum sensors~\cite{corgier_optimized_2025,pezze_quantum_2018}. For the Cramer-Rao SQL calculation, we differentiate the log-likelihood with respect to variations in \dPhi{} around a central parameter set, corresponding to the contrasts, atom numbers, and mean \dPhi{} extracted from a maximum likelihood fit to the full dataset in \cref{fig:phase_resolution}, and we further input the median measured atom numbers to the likelihood model. 
The dominant source of uncertainty in the Cramer-Rao SQL is the $\sim$ 7\% uncertainty in the atom-number calibration, and we calculate a standard deviation of \theoreticalSQLSingleShot{} per shot.
Over the whole dataset of \dataNumShotsSingle{} shots, this results in a lower bound of uncertainty in \dPhi{} of \theoreticalSQLWholeDataset{}.

\textbf{Monte-Carlo SQL:} We validate the unbinned maximum likelihood phase-extraction method and establish the SQL reference using Monte Carlo simulations that replicate the experimental sampling and noise budget. Synthetic shots include the measured contrasts, mean numbers of atoms and their fluctuations, and projection-noise-limited excitation readout, with the same estimator applied as in the real data analysis. These tests verify that the estimator is unbiased and that the observed phase variance is consistent with quantum projection noise under the measured atom-number statistics.

\begin{figure}[tb]
	\centering
    \includegraphics[width=\textwidth]{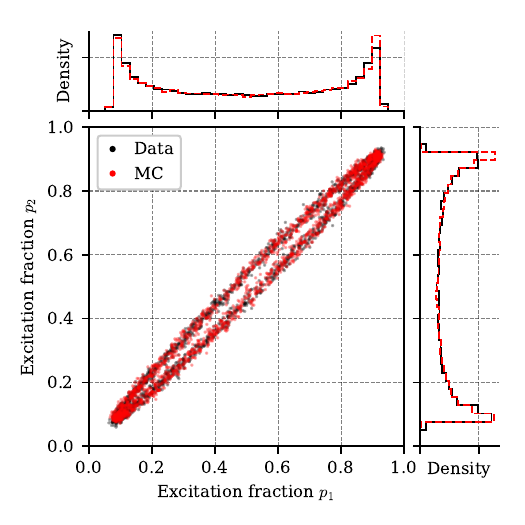}
\caption{%
    \textbf{A comparison between Monte-Carlo simulated data with experimental data.}
    Monte-Carlo simulated data contains only quantum projection noise (i.e.\ the Standard Quantum Limit) as a noise source. The ellipse plot in the $p_1$, $p_2$ plane shows \num{2000} randomly selected points, whereas the histograms summarise \num{30000} true and simulated shots.
}%
\label{fig:MC-comparison}
\end{figure}

We generated \numMCSims{} synthetic datasets with variation in atom number consistent with the uncertainty in mean from the absorption method described above, known shot-to-shot variation within datasets, and zero additional noise sources, shown in \cref{fig:MC-comparison}.
Each dataset consisted of \dataNumShotsSingle{}
simulated interferometer shots, each of which experiences contrasts of \useVal{qpn-contrast-top} and \useVal{qpn-contrast-bottom} for the two traps, and median numbers of atoms of \useVal{median-atoms-top} and \useVal{median-atoms-bottom} respectively, matching our real data. We include gaps in the simulated datasets to match the distribution of gaps in our true data, caused by various experimental calibrations / outages.
We verified that the recovered \dPhi{} values are unbiased within statistical uncertainty and that nominal \qty{68}{\percent} intervals have correct frequentist coverage.
By calculating an overlapping Allan deviation for each simulation run, then considering the distribution of Allan deviations across all generated datasets, we report \qty{68}{\percent} and \qty{95}{\percent} credible intervals for a differential interferometer limited only by atom shot noise (the SQL), shown in \cref{fig:phase_resolution}c. In contrast to the theoretical calculation, the Monte Carlo ensembles reproduce the full experimentally observed distributions of atom number, contrast, and projection-noise-limited excitation readout rather than only their mean values, and the resulting SQL reference is therefore a prediction analysed with the same estimator as the data.

Statistical compatibility with the SQL prediction was assessed using two complementary tests applied to the Allan deviation in log-space. A global test statistic comparing the measured values to Monte Carlo ensembles at each averaging time yields $p = 0.82$ for the HLN dataset and $p = 0.65$ for the LLN dataset, indicating no significant deviation from the Monte-Carlo SQL prediction. Additionally, the measured Allan deviation slopes ($s = -0.465$ for HLN, $s = -0.463$ for LLN) are consistent with the Monte Carlo SQL ensemble, which itself exhibits white-noise scaling ($s = -0.5$), with p = 0.45 and p = 0.43 respectively.

\textbf{Code availability:}
Experimental control python code and raw data for the results presented in this work are available at Ref.~\cite{zenodo_repository_raw}.
Analysed data and Jupyter notebooks to produce the figures in this paper are available at Ref.~\cite{zenodo_repository_analysis}.

\textbf{Gravitational Wave Detection Prospects:} In \cref{fig:gravitational-waves}, the outlined regions show the expected sensitivities of the indicated gravitational wave detectors at a signal-to-noise ratio SNR $\geq 8$ from the final stages of equal mass black hole mergers, after the binaries come within the innermost stable circular orbit, during one year of operation. The detector sensitivities are taken from~\cite{LIGOScientific:2014pky,Hild:2010id,LISA:2017pwj,Badurina:2021rgt}. The sensitivity of the AION-km detector incorporates an estimate of the possible mitigation of gravity gradient noise based on~\cite{Badurina:2022ngn,Carlton:2024lqy}. The cyan symbols are simulated GW signals from black hole mergers based on a model of hierarchical assembly of supermassive black holes starting from $100 M_\odot$ black hole seeds at redshift $z = 20$ that could be the remnants of collapsed first-generation Population III stars. This scenario is consistent with JWST and other observations of supermassive black holes~\cite{Ellis:2024nzv} and predicts $\sim 10^3$ AION-km detections per year. Scenarios with (an admixture of) heavier seeds and fewer detections are also consistent with the current supermassive black hole data. The violet dots are GW signals from mergers~\cite{Raidal:2024bmm} of a hypothetical population of primordial black holes computed assuming a primordial black hole (PBH) population that comprises 0.2\% of all the dark matter and has a log-normal mass distribution $\psi(M) \propto \exp\left(-\frac{\ln^2(M/M_c)}{2\sigma^2}\right)$ with a mean $M_c = 10 M_\odot$ and width $\sigma = 3$, which is consistent with the available constraints~\cite{Carr:2017jsz}. The orange dots are GW signals from a sample of stellar mass black hole mergers assuming a truncated power-law mass function in the mass range from $3 M_\odot$ to $60 M_\odot$ and a merger rate that follows the star formation rate, peaking around $z = 2.5$~\cite{Hutsi:2020sol,KAGRA:2021duu}. The exponent and the amplitude are fixed by fitting the data of~\cite{LIGOScientific:2014pky,VIRGO:2014yos,Aso:2013eba}.



\end{document}